\documentclass[3p,times,twocolumn]{elsarticle}

\usepackage{ecrc}

\volume{00}

\firstpage{1}

\journalname{Nuclear Physics B Proceedings Supplement}

\runauth{K. Maltman et al.}

\jid{nuphbp}

\jnltitlelogo{Nuclear Physics B Proceedings Supplement}

\usepackage{amssymb,amsmath}

\usepackage[figuresright]{rotating}
\usepackage{graphicx}
\usepackage{amsfonts, color, float, bbm, amsmath}

\begin{document}

\begin{frontmatter}




\title{Issues in Determining $\alpha_s$ from Hadronic $\tau$ Decay
and Electroproduction Data}


\author[barc1]{D. Boito}

\author[val,mun]{O. Cat\`a}

\author[sfsu]{M. Golterman}

\author[mj]{M. Jamin}

\author[kmy,kmcssm]{K. Maltman\corref{cor1}}

\author[sfsu]{J. Osborne}

\author[sfsu,barc2]{S. Peris}

\cortext[cor1]{Corresponding author}

\address[barc1]{Departament de F\'isica and IFAE,
Universitat Auton\`oma de Barcelona, E-08193 Bellaterra, Barcelona, Spain}

\address[val]{Departament de F\'isica Te\`orica and IFIC\\
Universitat de Val\`encia-CSIC, E-46071 Val\`encia, Spain}

\address[mun]{Ludwig-Maximilians-Universit\"at M\"unchen, Fakult\"at f\"ur 
Physik, Arnold Sommerfeld Center for Theoretical Physics, D–80333 M\"unchen, 
Germany}

\address[sfsu]{Department of Physics and Astronomy,
San Francisco State University, San Francisco, CA 94132, USA}

\address[mj]{ICREA, IFAE,  Universitat Auton\`oma de Barcelona,
E-08193 Bellaterra, Barcelona, Spain}

\address[kmy]{Dept. Math and Statistics, York University, Toronto, ON Canada
M3J 1P3}
\address[kmcssm]{CSSM, University of Adelaide, Adelaide, SA Australia 5005}

\address[barc2]{Departament de F\'isica, Universitat Auton\`oma de 
Barcelona, E-08193 Bellaterra, Barcelona, Spain}

\begin{abstract}
We discuss some key issues associated with duality-violating 
and non-perturbative OPE contributions to the theoretical representations 
of light quark current-current two-point functions and relevant
to precision determinations of $\alpha_s$ from hadronic $\tau$ decay 
and electroproduction cross-section data. We demonstrate that analyses 
with an explicit representation of duality-violating effects 
are required to bring theoretical errors in such extractions under control,
motivating the accompanying paper in these proceedings,
which presents the results of such an analysis.
\end{abstract}


\end{frontmatter}


\section{Introduction}
\label{intro}
Determinations of $\alpha_s$ based on hadronic $\tau$ decay or
electroproduction cross-section data rely on the fact
that, for any $s_0>0$ and any analytic $w(s)$, 
appropriate combinations, $\Pi (s)$, of the 
$J=0,1$ scalar components, $\Pi^{(J)}(s)$, 
of the relevant current-current two-point functions
satisfy the finite energy sum rule (FESR) relation
\begin{equation}
\int_{0}^{s_0}ds\, w(s)\, \rho (s)\ =\ {\frac{-1}{2\pi i}}
\, \oint_{\vert s\vert = s_0} ds \, w(s) \, \Pi (s)\, ,
\label{fesr}\end{equation}
with $\rho (s)= {\frac{1}{\pi}}\, Im\, \Pi (s)$ the spectral function
of $\Pi$. If the current in question is either of the flavor
$ud$ vector (V) or axial vector (A) currents, the LHS of Eq.~(\ref{fesr})
can be determined experimentally, for $s_0\le m_\tau^2$, using 
hadronic $\tau$ decay data~\cite{tsai,bnp}. For the 
electromagnetic (EM) current case, the LHS can similarly be determined 
using inclusive electroproduction cross-sections. A determination of 
$\alpha_s$ is then possible because, for large enough $s_0$, 
the OPE representation of the RHS is dominated by its dimension $D=0$ 
perturbative contribution. Additional contributions to the RHS result from
higher $D$, non-perturbative (NP) terms in the OPE 
representation of $\Pi$, as well as from any deviations
(``duality violations'' or DVs) between this representation and $\Pi$ itself. 
The precision with which $\alpha_s$ can be determined is affected 
by the accuracy with which these small, but not generally negligible,
contributions can be estimated.

With $Q^2=-s$, the OPE representation of $\Pi$ becomes
\begin{equation}
\left[\Pi (Q^2)\right]_{OPE}=\sum_{D=0,2,4,\cdots}{\frac{C_D}{Q^D}}\ .
\label{ope}\end{equation}
The $C_D$ depend logarithmically on $Q^2$. For $D=4$, the
coefficient functions multiplying
$\langle \alpha_s G^2\rangle$ and $\langle m_q\bar{q}q\rangle$
are known beyond leading order (see, e.g.,~\cite{bnp}). For $D>4$ we follow 
standard convention and treat $C_D$
as an effective constant. Corrections to this approximation are
suppressed by additional factors of $\alpha_s$ and accounted for
in an averaged sense if $C_D$ is fitted to data. To fit the $C_D$ 
it is convenient to employ polynomial weights
$w(y)=\sum_k w_k y^k$, with $y=s/s_0$, since OPE contributions to
the RHS of Eq.~(\ref{fesr}) with different $D$ then
scale differently with $s_0$. Such $w(y)$ yield $D\ge 6$ contributions
\begin{equation}
\sum_{k=2,3,\cdots} (-1)^k\, w_k\, {\frac{C_{2k+2}}{s_0^k}}\ .
\label{npcontribs}\end{equation}
A $w(y)$ of degree $N$ thus generates contributions, 
unsuppressed by additional powers of $\alpha_s$, up to $D=2N+2$.

``Pinched'' (``unpinched'') $w(s)$ are those with (without) a zero
at $s=s_0$. For $s_0\sim$ a few GeV$^2$, DV contributions 
to the RHS of Eq.~(\ref{fesr}) are known to be significant 
for unpinched $w(s)$~\cite{kmfesr,dv7}. Pinching in $w(s)$ significantly
reduces this effect~\cite{kmfesr}, compatible with the expectation that 
DV contributions to $\Pi$ will
be localized to the vicinity of the timelike point on the contour.

\section{$\tau$ Decay Analyses}\label{taustuff}
There are a number of recent $\alpha_s$ determinations based on 
non-strange hadronic $\tau$ decay
data [4-12],
all employing the 5-loop version of the
$D=0$ OPE Adler function series~\cite{bck08} and ALEPH~\cite{aleph05}
and/or OPAL~\cite{opal99} spectral data. 
Apart from Ref.~\cite{dv7},
all assume DVs can be neglected, though some, 
either fully or partly, check this assumption for 
self-consistency~\cite{davier08,my08}. Some~\cite{dv7,davier08,my08}, 
but not all, attempt to fit the relevant NP OPE 
coefficients{\footnote{Ref.~\cite{bj08} assumes values for the 
relevant $C_{6,8}$ which are also used implicitly in 
Refs.~\cite{menke09,cf}.}}.
In the Standard Model, with $R_{V/A}$
the ratios of flavor $ud$ V/A current-induced inclusive
$\tau$ decay widths to the corresponding electronic width, 
$y_\tau =s/m_\tau^2$, $S_{EW}$ a known short-distance EW
correction, $w_{(00)}(y)=(1-y)^2(1+2y)$ and $w_L(y)=y(1-y)^2$, 
one has~\cite{tsai}
\begin{eqnarray}
&&R_{V/A}=12\pi^2S_{EW}\, \vert V_{ud}\vert^2\, \int_0^{m_\tau^2}\, 
{\frac{ds}{m_\tau^2}}\nonumber\\
&&\qquad \left[ w_{00}(y_\tau )\rho^{(0+1)}_{V/A}(s)
\, -\,w_L(y_\tau )\rho_{V/A}^{(0)}(s)
\right] ,
\label{taubasic}\end{eqnarray}
with $\rho^{(J)}_{V/A}(s)$ the spectral function of $\Pi_{V/A}^{(J)}(s)$. 
$R_{V/A}$ are very accurately known experimentally. However,
the degree $3$ kinematic weights in Eq.~(\ref{taubasic}) 
produce NP contributions up to $D=8$ in the corresponding OPE representation,
making $R_{V/A}$ itself insufficient to determine
$\alpha_s$, even if DV contributions are assumed negligible for
the double-pinched weights $w_{(00)}$ and $w_L$. 

In Refs.~[6, 13-15]
an attempt was made to
fit $C_{6,8}$, and hence quantify the small NP OPE contributions to
$R_{V/A}$, by considering the ``$(km)$ spectral weight'' analogues, 
$R^{(km)}_{V/A}$, of $R_{V/A}$~\cite{ldp92}, 
obtained by reweighting the integrand 
in Eq.~(\ref{taubasic}) by $(1-y_\tau )^k y_\tau^m$, and using the 
$km=00,10,11,12$ and $13$ versions to fit the OPE parameters 
$\alpha_s$, $\langle \alpha_sG^2\rangle$, $C_6$ and $C_8$. 
This approach requires an additional implicit assumption.
The weights in $R_{V/A}^{(km)}$, $km=10,11,12,13$,
produce integrated, non-$\alpha_s$-suppressed OPE contributions up 
to $D=10,12,14,16$ respectively, with each new $R^{(km)}_{V/A}$ introducing
a new OPE parameter. The strategy is thus useful only if sufficiently 
many of the $C_{D>8}$ yield OPE contributions negligible
for all weights in the analysis. Refs.~[6, 13-15]
implicitly assume all $C_{D>8}$ can be neglected in this sense. If this
assumption fails, effects due to neglected, but non-negligible $D>8$
contributions will be absorbed into lower $D$ fitted parameters
and yield nominal OPE contributions scaling incorrectly with
$s_0$. The assumption can thus be tested by comparing 
experimental and fitted OPE versions of the 
$m_\tau^2\rightarrow s_0$, $y_\tau\rightarrow y$ generalizations of
the $R^{(km)}_{V/A}$. This test was performed for the V channel
fits of Ref.~\cite{davier08} (D08) in Ref.~\cite{my08} (MY08). 
While for the $(00)$ case (also tested 
in D08) the fit deviates from experiment 
only below $s_0\sim 2.5$ GeV$^2$, dramatic deviations are seen for the
other $(km)$ cases. Even more telling, the tests also fail for the 
doubly-pinched non-spectral $J=0+1$ weights $w_2(y)=1-2y+y^2$ 
and $w_3(y)=1-(3y/2)+(y^3/2)$ which, from Eq.~(\ref{npcontribs}),
test separately the $D=6$ ($w_2$) and $D=8$ ($w_3$) 
parts of the NP OPE contribution to $R_{V/A}$.
These problems could result from either non-negligible $D>8$, non-negligible 
residual DV contributions, or both. In either case, the results of the 
spectral weight analyses, including those of D08, are shown to be 
unreliable. In
assessing the reliability of other $\tau$-based $\alpha_s$ results in the 
literature, it should be borne in mind that Refs.~\cite{bck08,narison,pich11} 
did not themselves perform fits for the $C_{6,8}$ required to evaluate the 
$D=6,8$ contributions to $R_{V/A}$, but rather took these from one or more of
the above spectral weight analyses.

\begin{figure}[!ht]
\includegraphics[width=0.75\columnwidth,angle=270]
{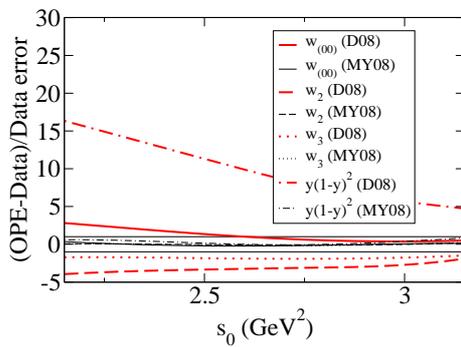}
\vspace{-0.95cm}
\caption{Comparison of the fit qualities for the OPE-only fits
of Refs.~\cite{davier08} and \cite{my08}, for a range of
degree $\le 3$ weights}
\label{vecfitqualities}
\end{figure}

In MY08, $\alpha_s$ and the $C_D$ were fitted using the $s_0$-dependence 
of various double-pinch-weighted spectral integrals.
Double-pinching was assumed sufficient to make residual DV contributions
negligible, but no assumptions about the $D>4$ OPE coefficients 
were made (i.e., all non-$\alpha_s$-suppressed NP OPE 
contributions were included). The resulting OPE fits yield excellent 
agreement between fitted OPE and experimental spectral integrals
for $s_0\gtrsim 2$ GeV$^2$, not only for the weights employed in the 
analysis, but also for the $km=00,10,11,12,13$ spectral weights.
Fig.~\ref{vecfitqualities} shows the 
MY08 (black curves) and D08 (red curves) V channel ``fit qualities'' 
(fitted OPE-experimental spectral integral differences 
scaled by the error on the latter) obtained from analyses of the 
same 2005 ALEPH data~\cite{aleph05},
for a range of degree $\le 3$ $w(s)$ (which
yield OPE integrals depending only the $D\le 8$ OPE parameters considered 
in D08). The results are obviously compatible with residual DVs being 
negligible and the problems in the D08 spectral weight analysis resulting
from a breakdown in the assumptions about $D>8$ contributions.

Two key issues remain concerning systematic errors in the MY08
approach. First, though it is consistent to neglect residual DV 
contributions for $s_0\gtrsim 2$ GeV$^2$, a quantitative
estimate of the uncertainty in $\alpha_s$ associated with
this neglect (which would require a model for DV contributions) is absent. 
The second issue concerns the use of the external charmonium 
sum rule input for $\langle \alpha_s G^2\rangle$~\cite{agginput}. MY08 were 
forced to employ this input because (i) neglect of DVs was found to
be inconsistent for the single-pinched, degree $1$, 
$w(y)=1-y$ FESR, which one might have hoped would allow a fit of 
$\langle \alpha_sG^2\rangle$ using $\tau$ data alone; 
(ii) doubly-pinched weights, with better suppression of DV contributions, 
produce OPE contributions up to at least $D=6$; and (iii) the window of $s_0$ 
for which the neglect of DVs in doubly-pinched FESRs is self-consistent 
is insufficiently broad to allow simultaneous extraction of 
$\alpha_s$, $\langle \alpha_s G^2\rangle$ and one of the higher $C_D$. 
The use of external input is potentially problematic because of the 
renormalon ambiguity in the definition of $\langle \alpha_s G^2\rangle$.
This might require different values for the {\it effective} 
$\langle \alpha_s G^2\rangle$ obtained by analyzing different correlators 
treated with different $D=0$ series truncation orders. If so, the
$D=4$-induced error on $\alpha_s$ might be significantly larger than 
that generated using only the uncertainty on $\langle\alpha_s G^2\rangle$ 
from the charmonium sum rule analysis. We now elaborate on,
and illustrate, some of the points just made.

\begin{figure}[!ht]
\includegraphics[width=0.75\columnwidth,angle=270]
{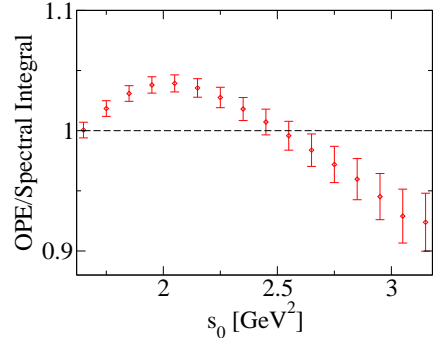}
\vspace{-0.92cm}
\caption{Duality violations for the V channel $w=1$ FESR}
\label{vecdvsweq1}
\end{figure}

\begin{figure}[!ht]
\includegraphics[width=0.75\columnwidth,angle=270]
{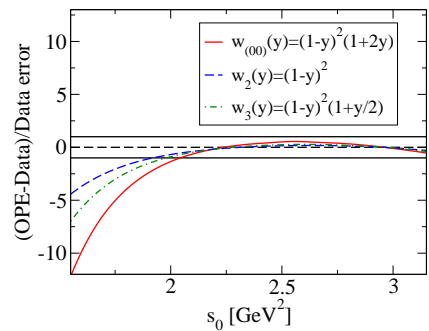}
\vspace{-0.92cm}
\caption{MY08 $w_{(00)}$, $w_2$ and $w_3$ V channel fit qualities}
\label{vecdvsdblpinch}
\end{figure}

\begin{figure}[!ht]
\includegraphics[width=0.75\columnwidth,angle=270]
{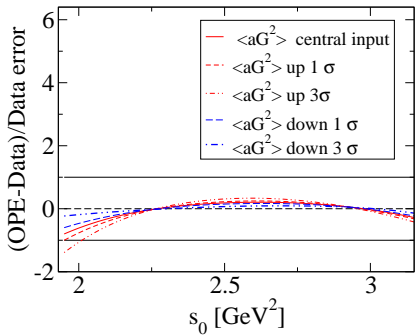}
\vspace{-0.95cm}
\caption{MY08 $w_2$ V channel fit qualities
with input $\langle \alpha_s G^2\rangle$ from $3\sigma$
below to $3\sigma$ above the charmonium sum rule~\cite{agginput}
central value.}
\label{aggfitqualities}
\end{figure}

First, regarding DVs, it is clear that, even for $s\sim 3$ GeV$^2$, 
significant DVs are evident in the experimental spectral 
functions~\cite{aleph05,opal99}.
Considering integrals over the spectral functions does not necessarily
improve the situation, as demonstrated in Fig.~\ref{vecdvsweq1}, which 
shows the ratio of V channel OPE to spectral integrals for the unpinched
weight $w(s)=1$. The error bars reflect the spectral integral
errors, obtained from the experimental covariance matrix. 
Fig.~\ref{vecdvsdblpinch}, which shows the fit qualities corresponding
to the MY08 analysis for the $J=0+1$ doubly-pinched 
$w_{(00)}$, $w_2$ and $w_3$ FESRs, demonstrates that, while residual 
DVs may be small above $s_0\gtrsim 2$ GeV$^2$, they
turn on rapidly below this point, making an estimate of the
residual contributions in the MY08 fit window highly desirable.
Finally, Fig.~\ref{aggfitqualities} illustrates the inability to
obtain well-constrained fits to more than two OPE parameters in
the limited fit window above $s_0\sim 2$ GeV$^2$. The figure
shows the fit qualities for the optimized $w_2$ V channel fits for 
input $\langle \alpha_s G^2\rangle$ ranging from $3\sigma$
below to $3\sigma$ above the central charmonium sum rule value.
It is obvious that all input $\langle \alpha_s G^2\rangle$ in this range
produce essentially equally good optimized fits, and hence that
the $\tau$ data alone cannot be used to successfully
constrain $\langle \alpha_s G^2\rangle$. The strong anti-correlation 
between output $\alpha_s$ and input $\langle \alpha_s G^2\rangle$
means that, if one were maximally conservative, and attempted to rely 
on the $\tau$ data alone, sticking to the doubly-pinched weights for 
which residual DVs are safely small, one would be left with a
contribution to the error on $\alpha_s$ easily three times the
component quoted in MY08 associated with the charmonium sum 
rule error on $\langle \alpha_s G^2\rangle$. Such an expanded assessment
of the $D=4$ input uncertainty yields a contribution $\sim 0.0020$ 
to the uncertainty on $\alpha_s^{n_f=5}(M_Z^2)$, already larger than
the error quoted in any of the recent analyses not including DVs.

To summarize the $\tau$ situation:
(i) residual DVs can be plausibly
neglected in the kinematically accessible region 
only for doubly-pinched weights, and, for these, only in the window 
$s_0\gtrsim 2$ GeV$^2$;
(ii) using $\tau$ data alone, such an $s_0$ window is insufficiently wide 
to allow the NP OPE coefficients needed for a precision determination of 
$\alpha_s$ to be fit with sufficient accuracy to achieve the 
precision claimed by recent analyses which neglect DV effects; (iii) this
situation can be improved by extending the 
fit window to lower $s_0$ and/or including additional 
degree $0$ and/or $1$ weights (which introduce no new OPE parameters, 
but are, unavoidably, at most singly pinched); (iv) for either of these
improvement options, residual DVs will no longer be negligible;
and (v) {\it a reliable improved analysis thus necessarily requires use of 
some model to account for DV effects.} The accompanying paper (see M.
Jamin, these proceedings) reports on the results of such an analysis, obtained 
using the DV model of Refs.~\cite{cgp}. 

\section{Electroproduction-cross-section-based analyses}\label{EMstuff}

The spectral function, $\rho_{EM}(s)$, of the EM current scalar two-point
function, $\Pi_{EM}(s)$, is related to the inclusive bare electroproduction 
cross-section by
\begin{equation}
\rho_{EM}(s)\, =\, {\frac{s\, \sigma_{bare}(s)}{16\pi^3\alpha_{EM} (0)^2}}\, ,
\label{rhoemcrosssectionreln}\end{equation}
allowing $\alpha_s$ to be determined from FESRs 
based on inclusive electroproduction data. An advantage of this approach
over that based on $\tau$ decay data is that no kinematic bound 
exists on the upper limit, $s_0$, of the spectral integral 
in Eq.~(\ref{fesr}). Since the integrated $D=0$ OPE contribution
to the RHS of this equation grows roughly linearly with $s_0$, while both 
integrated NP OPE and integrated DV contributions decrease with increasing
$s_0$, the theoretical representation is more and more dominated
by its perturbative, $\alpha_s$-dependent contribution as $s_0$ increases.
In addition, the $I=1$ part of $\Pi_{EM}$ is related by CVC to 
$\Pi^{(1)}_V$, allowing the $\tau$ data to be used to
(i) estimate the small $I=1$ higher $D$ NP OPE and
DV contributions to the RHSs of the EM FESRs and (ii) identify those $s_0$ 
for which such contributions can be safely neglected. 
Presently the experimental situation is complicated by 
(i) discrepancies between the BaBar and KLOE $\pi\pi$ cross-section 
data~\cite{babarkloepipi} and (ii) discrepancies much larger than 
typically associated with isospin-breaking effects between 
preliminary BaBar~\cite{dr07} and SND~\cite{sndphipsi} 
$\pi^+\pi^-\pi^0\pi^0$ cross-sections and expectations based on 
experimental $\tau\rightarrow 4\pi\nu_\tau$ distributions and CVC.
Some very preliminary exploratory results were presented in the
conference talk, and work on this analysis is ongoing.

\vspace{3ex}
\noindent {\bf Acknowledgments}

{\it We thank Martin Beneke, Claude Bernard, Andreas H\"ocker, 
Manel Martinez, and Ramon Miquel for useful discussions and
Sven Menke for significant help with understanding the
OPAL spectral-function data. MG thanks IFAE and the Department of 
Physics at UAB, and OC and KM thank the 
Department of Physics and Astronomy at SFSU for hospitality.
DB, MJ and SP are supported by CICYTFEDER-FPA2008-01430, SGR2005-00916, 
the Spanish Consolider-Ingenio 2010 Program CPAN (CSD2007-00042). 
SP is also supported by  a fellowship from the Programa de Movilidad
PR2010-0284. OC is supported in part by MICINN (Spain) under Grant 
FPA2007-60323, by the 
Spanish Consolider Ingenio 2010 Program CPAN (CSD2007-00042) and by the 
DFG cluster of excellence ``Origin and Structure of the Universe.''
MG and JO are supported in part by the US Department of Energy, and
KM is supported by a grant from the Natural Sciences and
Engineering Research Council of Canada.}



\begin{thebibliography}{00}
\bibitem{tsai}Y.-S. Tsai, {\it Phys. Rev.} {\bf D4} (1971) 2821.
\bibitem{bnp}E. Braaten, S. Narison, A. Pich, {\it Nucl. Phys.} {\bf B373}
(1992) 581.
\bibitem{kmfesr}K. Maltman, {\it Phys. Lett.} {\bf B440} (1998) 367;
C. A. Dominguez, K. Schilcher, {\it Phys. Lett.} {\bf B448} (1999) 93.
\bibitem{dv7}D. Boito {\it et al.}, {\it Phys. Rev.} {\bf D84} (2011) 113006.
\bibitem{bck08}P.A. Baikov, K.G. Chetyrkin, J.H. K\"uhn,
{\it Phys. Rev. Lett.} {\bf 101} (2008) 012002.
\bibitem{davier08}M. Davier {\it et al.}, {\it Eur. Phys. J.} {\bf C56}
(2008) 305.
\bibitem{bj08}M. Beneke, M. Jamin, {\it JHEP} {\bf 0809} (2008) 044.
\bibitem{my08}K. Maltman, T. Yavin, {\it Phys. Rev.} {\bf D78} (2008) 094020.
\bibitem{menke09}S. Menke, arXiv:0904.1796 [hep-ph].
\bibitem{cf}I. Caprini, J. Fischer, {\it Phys. Rev.} {\bf D84} (2011)
054019
\bibitem{narison}S. Narison, {\it Phys. Lett.} {\bf B673} (2009 30.
\bibitem{pich11}A. Pich, arXiv:1107.1123 [hep-ph].
\bibitem{aleph05}S. Schael {\it et al.} (ALEPH),
{\it Phys. Rep.} {\bf 421} (2005) 191.
\bibitem{opal99}K. Anderson {\it et al.} (OPAL),
{\it Eur. Phys. J.} {\bf C7} (1999) 571.
\bibitem{aleph98}R. Barate {\it et al.} (ALEPH),
{\it Eur. Phys. J.} {\bf C4} (1998) 409.
\bibitem{ldp92}F. Le Diberder, A. Pich, {\it Phys. Lett.} {B289} (1992) 
165.
\bibitem{agginput}B.L. Ioffe, K.N. Zyablyuk, {\it Eur. Phys. J.} {\bf C27}
(2003) 229.
\bibitem{cgp}O. Cat\`a, M. Golterman, S. Peris, {\it Phys. Rev.}
{\bf D77} (2008) 093006; {\it ibid.} {\bf D79} (2009) 053002.
\bibitem{babarkloepipi}B. Aubert {\it et al.} (BaBar), {\it Phys. Rev.
Lett.} {\bf 103} (2009) 231801; F. Ambrosino {\it et al.} (KLOE),
{\it Phys. Lett.} {\bf B700} (2011) 102 and P. Lukin, these proceedings.
\bibitem{dr07}V. Druzhinin, arXiv:0710.3455 [hep-ex].
\bibitem{sndphipsi}See S.I. Serednyakov, these proceedings.

\end{thebibliography}



\end{document}